# Tunnel Junctions on As-Grown Superconducting MgB$_2$ Thin Films

K. Ueda, M. Naito

*Abstract*—We demonstrate that good superconductor/insulator/ normal-metal tunnel junctions can be fabricated on as-grown superconducting MgB$_2$ thin films. The as-grown films were prepared by coevaporation at low growth temperatures of around 280˚C. The insulating barrier was formed by an Mg overlayer, which is subsequently oxidized in air. The tunneling spectra for Au/MgO$_x$/MgB$_2$ junctions showed a reproducible and well-defined superconducting gap ($\Delta$ = ~ 2.5 meV). The resultant 2$\Delta$/ $k_B T_c$ was significantly smaller than the predicted BCS value of 3.52.

## I. INTRODUCTION

The discovery of superconductivity at 39 K in MgB$_2$ [1] has generated great scientific and technological interest. As compared with cuprates, MgB$_2$ may be more suitable for fabricating good tunnel junctions because it has less anisotropy, fewer material complexities, fewer interface problems, and a longer coherence length ($\xi$ = ~ 5 nm) [2]. The importance of tunnel junctions is twofold: they are used for electronics applications and for investigating the pairing interaction. As regards the former aspect, good MgB$_2$ junctions are desired despite having a lower $T_c$ than cuprates since reliable and reproducible processes for fabricating good cuprate Josephson junctions have yet to be established even after it is 15 years since their discovery [3], [4]. In terms of the latter aspect, evidences have been accumulating that MgB$_2$ is essentially a conventional BCS superconductor. For example, the B and/or Mg isotope effects indicate that B atom vibrations are significantly involved in the superconductivity of MgB$_2$ [5], [6]. Photoemission spectroscopy [7], scanning tunneling spectroscopy [8], and NMR [9] have also shown that the superconducting gap is s-wave like. However, there is still considerable controversy with regard to the magnitude and character of the superconducting gap [8], [10]-[16]. Some early results indicated an extremely small gap value of less than 3 meV corresponding to 2$\Delta$/ $k_B T_c$ < 2.0 [12]-[16]. Even the existence of multiple gaps has been predicted by several groups [11], [12], [16]-[18]. Good tunnel junctions may settle this controversy. Furthermore good tunnel junctions are required to extract the Eliashberg function $\alpha^2 F(\omega)$, and thereby allow us to determine which phonon is crucial in terms of $T_c$ ~ 39 K.

High-quality MgB$_2$ films are a prerequisite for reliable fabrication of tunnel junctions. There have already been many reports on the synthesis of superconducting MgB$_2$ films. However, most require high-temperature post-annealing processes (600 ~ 900°C) [19]-[27], which are unsuitable for fabricating junctions because of the reaction and mixing at the interface at high temperatures. As-grown synthesis based on low-temperature processes is a desirable way of fabricating junctions. Recently, we have succeeded in preparing as-grown superconducting films at low temperatures (~300 ºC) by coevaporation in a UHV chamber [28]. In this article, we describe the as-grown synthesis of MgB$_2$ films and preliminary attempts at the fabrication of superconductor/insulator /normal-metal (SIN) tunnel junctions using as-grown MgB$_2$ thin films.

## II. EXPERIMENTAL

MgB$_2$ films were grown in a customer-designed MBE chamber (basal pressure of < 2 × 10$^{-9}$ Torr) from pure metal sources using multiple electron beam evaporators [29], [30]. Films were prepared mostly on Al$_2$O$_3$ c-plane substrates at growth temperatures lower than 300°C. The evaporation beam flux of each element was controlled by electron impact emission spectrometry (EIES) via feedback loops to the electron guns. The flux ratio of Mg to B was set at three times the nominal flux. The growth rate was 1.5 - 2 Å/s, and the film thickness of MgB$_2$ was typically 1000 Å. The structure and crystallinity were characterized by reflection high-energy electron diffraction (RHEED) and X-ray diffraction (XRD; 2$\theta$ $-\theta$ scan).

We fabricated SIN junctions using these MgB$_2$ films. As an insulating barrier, a 10 ~ 30 Å thick Mg overlayer was deposited in-situ, and subsequently oxidized in air for 1 ~ 30 days. Then counterelectrodes of normal metal (Au or Ag) were deposited. The junction area was defined by painting it with polystyrene Q-dope and was typically ~ 500 × 100 µm.

### III. RESULTS AND DISCUSSION

As-grown superconducting $MgB_2$ films can be prepared at $T_s$ (substrate temperature) values below 300°C as reported in our previous paper [28]. There is one trend within this growth temperature range, namely that a higher growth temperature leads to a higher superconducting transition temperature $T_c^{zero}$. At growth temperatures above 280°C, however, the Mg sticking coefficient decreases dramatically, and hence it is not easy to obtain the desired $MgB_2$ stoichiometry reproducibly, even when using beam fluxes with significantly (e.g. 10 times) enriched Mg. Therefore we prepared all the films reported in this paper at $T_s = 280°C$ in order to obtain the reproducible results. Furthermore, we learned from our recent systematic investigation [31] that the presence of residual oxygen during growth, even with $P_{O2}$ of as low as $1 \times 10^{-8}$ Torr, greatly degrades the resultant film quality. Therefore the growth was performed in $P_{O2}$ at below $4 \times 10^{-10}$ Torr, which also improved the reproducibility considerably. The films were c-axis orientated, and had a slightly depressed superconducting transition temperature ($T_c^{zero}$ of 33 - 35 K), a room-temperature resistivity of 30 – 40 $\mu\Omega$ cm, and an RRR of ~ 1.3. A typical $\rho$-T curve is shown in Fig.1. The as-grown films did not show any noticeable deterioration after one year in a dry box.

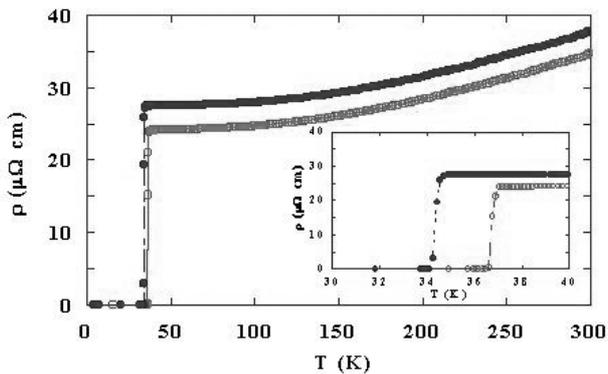

Fig. 1. Temperature dependence of resistivity of the as-grown film (●) and the post annealed (in-situ annealed) film (○). The inset shows an enlargement of $\rho$-T curves around $T_c$.

To improve the superconducting properties further, we attempted in-situ post-annealing for these as-grown films. This was performed just after growth at 480°C for 10 minutes in a vacuum while exposing the films to Mg flux (~ 4.5 Å/sec) to avoid any loss of Mg. Based on the intensities of the X-ray diffraction (XRD) peaks, there was no meaningful difference in the crystallinity of the post-annealed and the as-grown films. However, the $T_c^{zero}$ improved to 36 – 37 K, which is 2 - 3 K higher than that of the as-grown film. The resistivity also fell slightly as shown in Fig. 1. This indicates that the improved superconductivity resulting from the post-annealing may be due to a partial elimination of an unfavorable property in the grain boundaries.

AFM images (1 μm × 1 μm) of the as-grown and in-situ annealed films are compared in Fig. 2(a) and (b). The grain size was almost the same in both films (200 – 400 Å), namely in-situ annealing at 480°C did not enlarge the grain size. This statement is consistent with the above XRD results. The surfaces of both films were fairly smooth, and the root-mean-square roughness ($R_{MS}$) and the average roughness ($R_a$) were 22.3 Å and 17.7Å for the as-grown films and 17.4 Å and 13.7Å for the post-annealed films.

Next we provide the tunneling results for these as-grown and post-annealed films. Figure 3(a) shows the temperature dependence (1.4 ~ 41.1 K) of dI/dV for one of the best $Au/MgO_x/MgB_2$ junctions. This specific junction was fabricated with a 30 Å Mg overlayer, which was subsequently oxidized in air. The junction showed typical SIN characteristics with a well-defined superconducting gap ($\Delta$) of ~2.5 meV, and the gap closed at around $T_c^{zero}$ (~33 K) of this film. The $2\Delta/k_BT_c$ value was ~1.5, which is significantly smaller than the value of 3.52 predicted by BCS. Similar behavior was observed in Au/MgO$_x$/MgB$_2$ junctions with Mg overlayers with different thicknesses (10-30 Å). In fact, the junction resistance was not exponentially dependent on the thickness of the Mg overlayer. This indicates that the MgO$_x$ barrier was not ideally formed. We think that this is due to insufficient oxidation of the Mg overlayer or the incorporation of Mg in the Mg-deficient surface of the MgB$_2$ films from the overlayer.

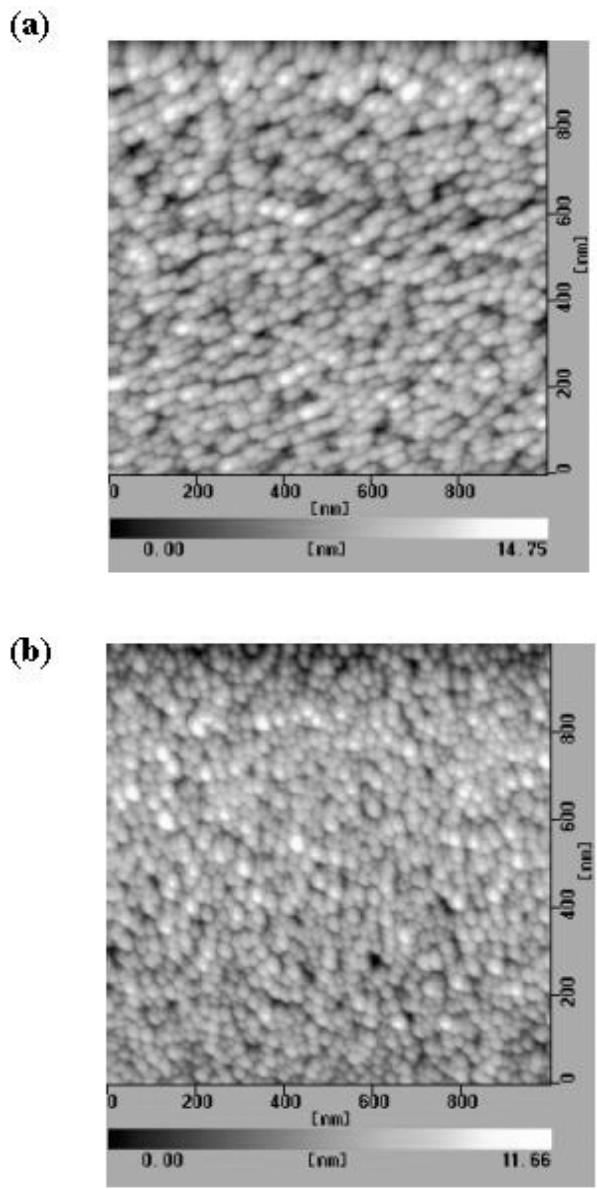

Fig. 2. AFM images of (a) the as-grown film and (b) the post- annealed (in-situ annealed) films.

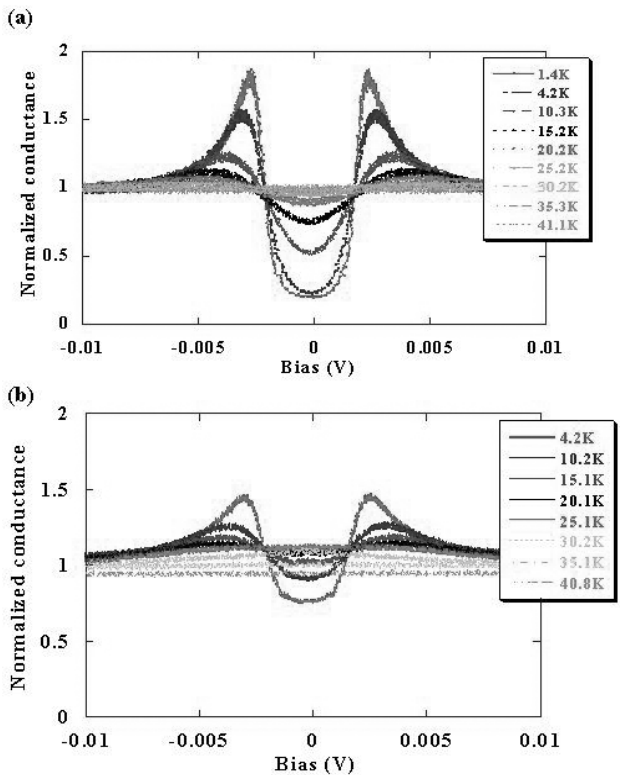

Fig. 3. Temperature dependence (1.4 ~ 41.1 K) of tunneling spectra of (a) the Au/MgO$_x$(Mg:30Å)/MgB$_2$ and (b) the Au/MgB$_2$ (in-situ annealed films) junction.

We also fabricated tunnel junctions on post-annealed films. The post-annealed films had a higher $T_c$ (~ 37 K) than the as-grown films, and hence we assumed that they *would* give better results. However, as shown in Fig. 3(b), the resultant tunnel spectra were inferior to those on as-grown films. Although the gap magnitude was almost the same as that observed for the as-grown film, the zero-bias conductance was much higher and the enhancement of the density of states at the gap edge was much weaker. The reason for this poor behavior is most likely due to the presence of a deteriorated surface layer [31]. This speculation is based on our investigation of the superconducting properties of ultra-thin films. In Fig. 4, the $T_c$ values are plotted as a function of film thickness for both as-grown films and post-annealed films. For films thicker than 100Å, the post-annealed films had a slightly higher $T_c$ than the as-grown films. For films thinner than 50Å, however, the situation was reversed. In fact, post-annealed films with a 50Å thickness are insulating whereas as-grown films with the same thickness are superconducting with a $T_c$ of ~ 9 K. This indicates that a 50Å surface layer, at least, is not superconducting. We think that this deteriorated layer is due to a significant Mg deficiency at the surface caused by the reevaporation of Mg during post-annealing [31]. The post-annealing process improved the $T_c$ of the interior portion of films by a few K, but degraded the superconductivity at the surface. This clearly indicates that post-annealed films are less suitable for fabricating junctions than as-grown films in spite of their higher $T_c$.

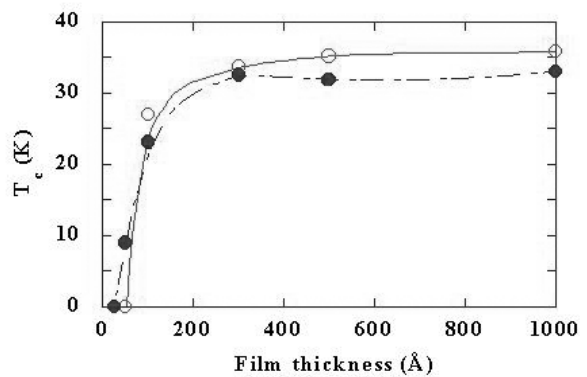

Fig. 4. Film thickness dependence of $T_c$ of as-grown films (●) and post-annealed films (○).

Finally we comment on the small superconducting gap (~ 2.5 meV) of MgB$_2$. First, we should mention that many groups have observed a similar gap value although the reason for the small gap is not well established [8], [12]-[16]. If we consider this small gap value to be a true bulk property, one possible explanation may be two-gap superconductivity with two distinct gap values (small gap $\Delta_S$ and large gap $\Delta_L$). The observed gap value of ~ 2.5 meV corresponds to $\Delta_S$. This possibility was initially proposed by Liu *et al.* on the basis of band calculations, which predicted two types of Fermi surface for MgB$_2$: a large two-dimensional surface and a small three-dimensional surface. Two-gap superconductivity has been supported experimentally by Tsuda *et al.* using high-resolution photoemission spectroscopy [18] and by Giubileo *et al.* using scanning tunneling microscopy [11]. In both of cases they claimed the existence of two distinct gaps (small gap $\Delta_S$ ~ 1.7 – 3.8 meV and large gap $\Delta_L$ ~ 5.4 – 7.8 meV). However, one important note of caution has to be sounded. Tunneling spectroscopy is a surface-sensitive measurement, and it essentially probes the surface of superconductors, whose properties may differ significantly from the true bulk properties. With regard to this problem, we can learn a good lesson from high-$T_c$ cuprates, for which we have recently demonstrated that the surface and bulk properties are totally different [3], [4]. Anyway, considerable caution must be exercised when investigating the nature and magnitude of the intrinsic superconducting gap of $MgB_2$.

IV. SUMMARY

Good SIN tunnel junctions were fabricated on as-grown superconducting $MgB_2$ thin films. The tunneling spectra for Au/MgO$_x$/MgB$_2$ junctions showed a reproducible and well-defined superconducting gap ($\Delta = \sim 2.5$ meV). The resultant $2\Delta/k_BT_c$ was $\sim 1.5$, which is significantly smaller than the predicted BCS value of 3.52. The reason for this discrepancy is not well understood, although a deterioration in the superconductivity at the surface is one possibility. Furthermore, we compared the tunneling spectra of as-grown and post-annealed films. The results indicated that post-annealed films were less suitable for fabricating junctions than as-grown films in spite of their higher $T_c$. This is because of the presence of thicker deteriorated layers in post-annealed films.

ACKNOWLEDGMENT

The authors thank Dr. S. Karimoto, Dr. H. Yamamoto, Dr. H. Sato and Dr. J. Nitta for fruitful discussions, and Dr. H. Takayanagi and Dr. S. Ishihara for their support and encouragement throughout the course of this study.